\documentclass[11pt]{article}
\usepackage{moriond,epsfig,url}

\newcommand{\be}{\begin{equation}}
\newcommand{\ee}{\end{equation}}
\newcommand{\bea}{\begin{eqnarray}}
\newcommand{\eea}{\end{eqnarray}}
\newcommand{\bi}{\begin{itemize}}
\newcommand{\ei}{\end{itemize}}
\newcommand{\ben}{\begin{enumerate}}
\newcommand{\een}{\end{enumerate}}

\newcommand{\lp}{\left(}
\newcommand{\rp}{\right)}

\begin{document}
\begin{flushright}
FR-PHENO-2010-19\\
Edinburgh 2010/09\\
CP3-10-16\\
\end{flushright}
\vspace*{1cm}
\title{COMBINED PDF AND STRONG COUPLING UNCERTAINTIES AT THE
LHC WITH NNPDF2.0}

\author{ The NNPDF Collaboration:\\
 Maria~Ubiali$^{1,5}$, Richard~D.~Ball$^{1}$,
 Luigi~Del~Debbio$^1$, Stefano~Forte$^2$,\\ Alberto~Guffanti$^3$, 
Jos\'e~I.~Latorre$^4$ and 
Juan~Rojo$^2$}

\address{\it ~$^1$ School of Physics and Astronomy, University of Edinburgh,\\
JCMB, KB, Mayfield Rd, Edinburgh EH9 3JZ, Scotland\\
~$^2$ Dipartimento di Fisica, Universit\`a di Milano and
INFN, Sezione di Milano,\\ Via Celoria 16, I-20133 Milano, Italy\\
~$^3$  Physikalisches Institut, Albert-Ludwigs-Universit\"at Freiburg
\\ Hermann-Herder-Stra\ss e 3, D-79104 Freiburg i. B., Germany  \\
~$^4$ Departament d'Estructura i Constituents de la Mat\`eria, 
Universitat de Barcelona,\\ Diagonal 647, E-08028 Barcelona, Spain\\
~$^5$ Center for Particle Physics Phenomenology CP3, Universit\`e Catholique de Louvain \\
Chemin du Cyclotron, 1348 Louvain-la-Neuve, Belgium}

\maketitle\abstracts{We present predictions for relevant
LHC observables obtained with the NNPDF2.0 set.
We compute the
combined PDF+$\alpha_s$ uncertainties on these observables,
and show that combining errors in quadrature yields an
excellent approximation to exact error propagation. We then
compare the NNPDF2.0 results to the other global PDF fits using
a common value of $\alpha_s$. At LHC 7 TeV, reasonable
agreement, both in central values and in uncertainties,
is found for NNPDF2.0, CTEQ6.6 and MSTW08.}

\section{Combined PDF+$\alpha_s$ uncertainties for LHC
observables}

The determination of the theoretical accuracy in the predictions 
for LHC observables is one of the most important tasks  now that the LHC is producing collisions 
at $\sqrt{s}=$7 TeV. QCD uncertainties coming from Parton Distribution Functions (PDFs)
and from the strong coupling constant $\alpha_s\lp M_Z\rp$ are among the 
dominant sources of theoretical uncertainties for most relevant LHC cross sections.

In this contribution we present predictions for important LHC observables
based on the NNPDF2.0 global PDF analysis~\cite{Ball:2010de}. First we will
discuss the results for the combined PDF+$\alpha_s$ uncertainty
on several LHC observables, and then we compare the NNPDF2.0 predictions with those of the other two global
analyses, MSTW2008 and CTEQ6.6. For the latter comparison
we use the sets with varying $\alpha_s$ recently presented
by these two groups~\cite{Martin:2009bu,Lai:2010nw} in order to use consistently a common value of $\alpha_s$.
The observables have been computed with the MCFM program~\cite{mcfm}.
We point out that predictions from previous NNPDF sets~\cite{DelDebbio:2007ee,Ball:2008by,Ball:2009mk,Ball:2009qv} are consistent with the NNPDF2.0 results, albeit with larger PDF uncertainties due to the reduced dataset used there.

First of all we present results for several LHC observables at 7 TeV computed
with the NNPDF2.0 PDF set: $W^{+}$ and $Z^0$ production,
$t\bar{t}$ production and Higgs production in gluon--fusion
for $m_H=120$ GeV. 
We compute predictions for various values of
$\alpha_s$ in order to determine the combined PDF+$\alpha_s$
uncertainties for these observables. Our choice
for the reference value of $\alpha_s$ and its uncertainty is
$\alpha_s\lp M_Z\rp= 0.119 \pm 0.002$, where the
uncertainty is to be interpreted as a 68\% C.L. The combined PDF+$\alpha_s$
uncertainty is computed both adding in quadrature the two
uncertainties and using exact error propagation, following
the methods presented in Refs.~\cite{Binoth:2010ra,Demartin:2010er}. 

\begin{figure}[ht]
\begin{center}
\epsfig{width=0.45\textwidth,figure=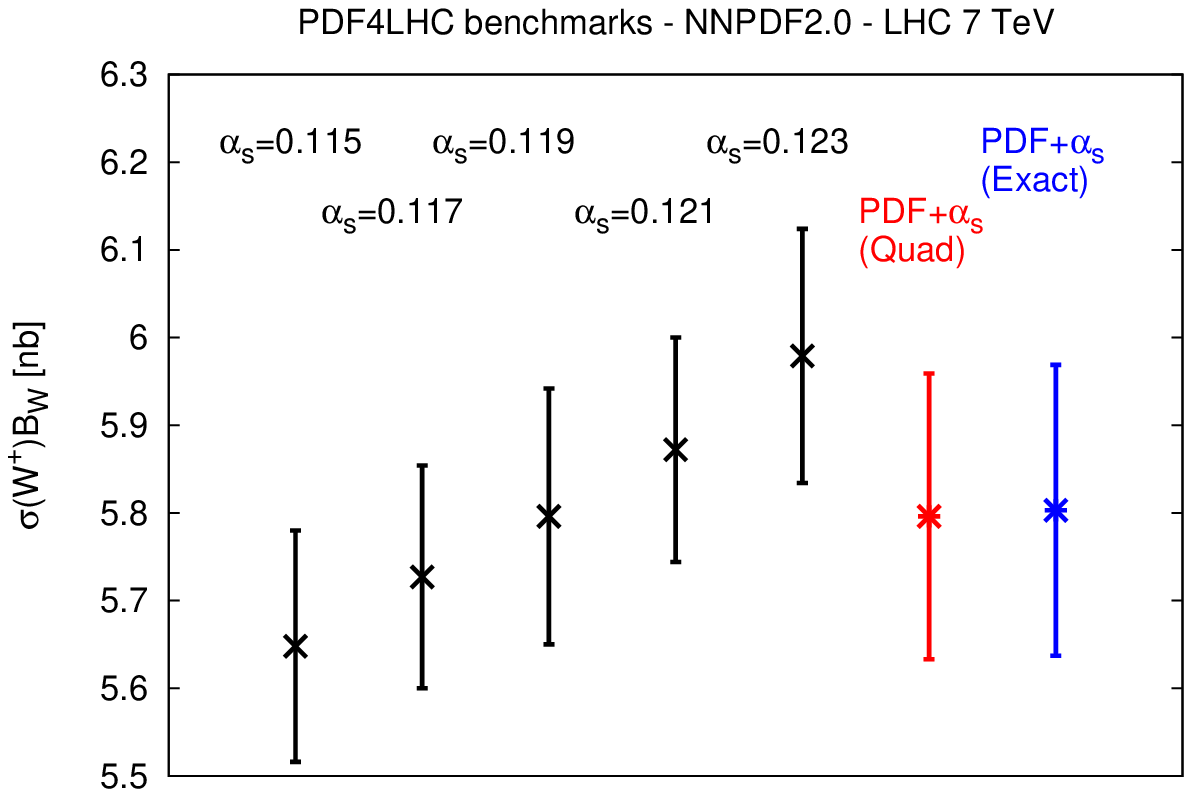}
\epsfig{width=0.45\textwidth,figure=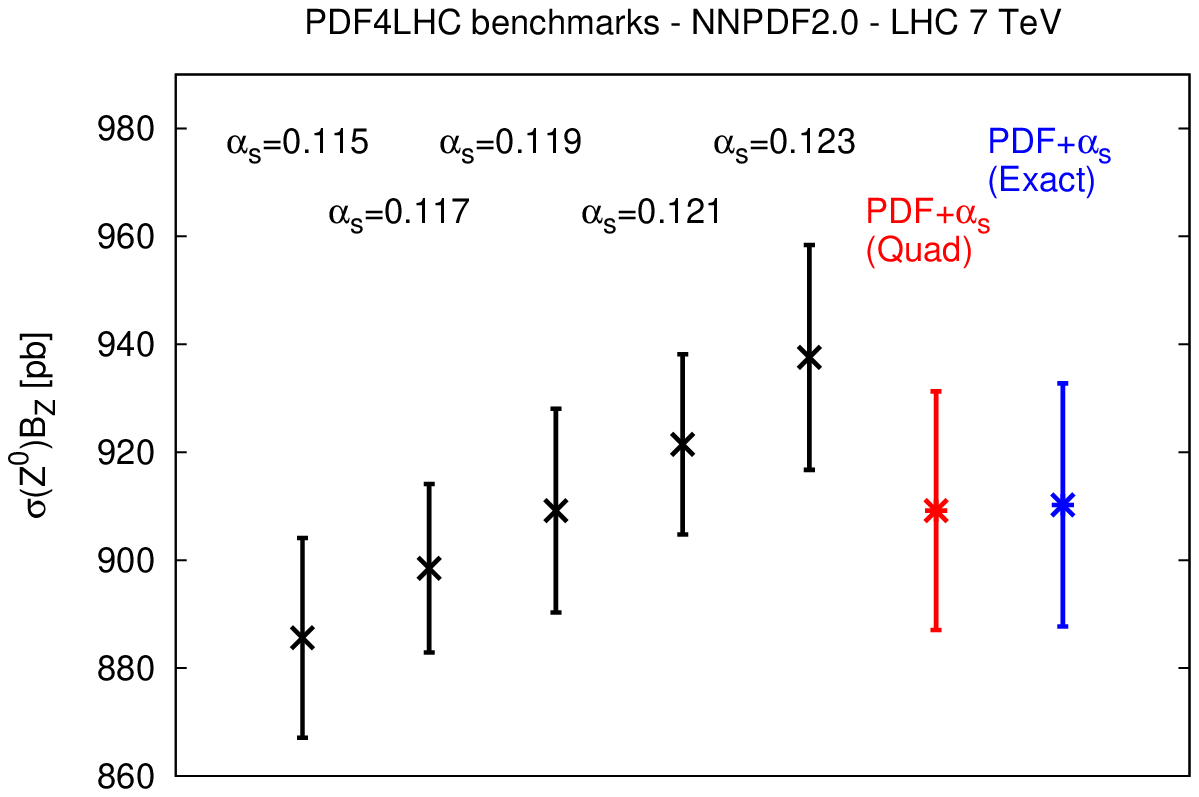}
\epsfig{width=0.45\textwidth,figure=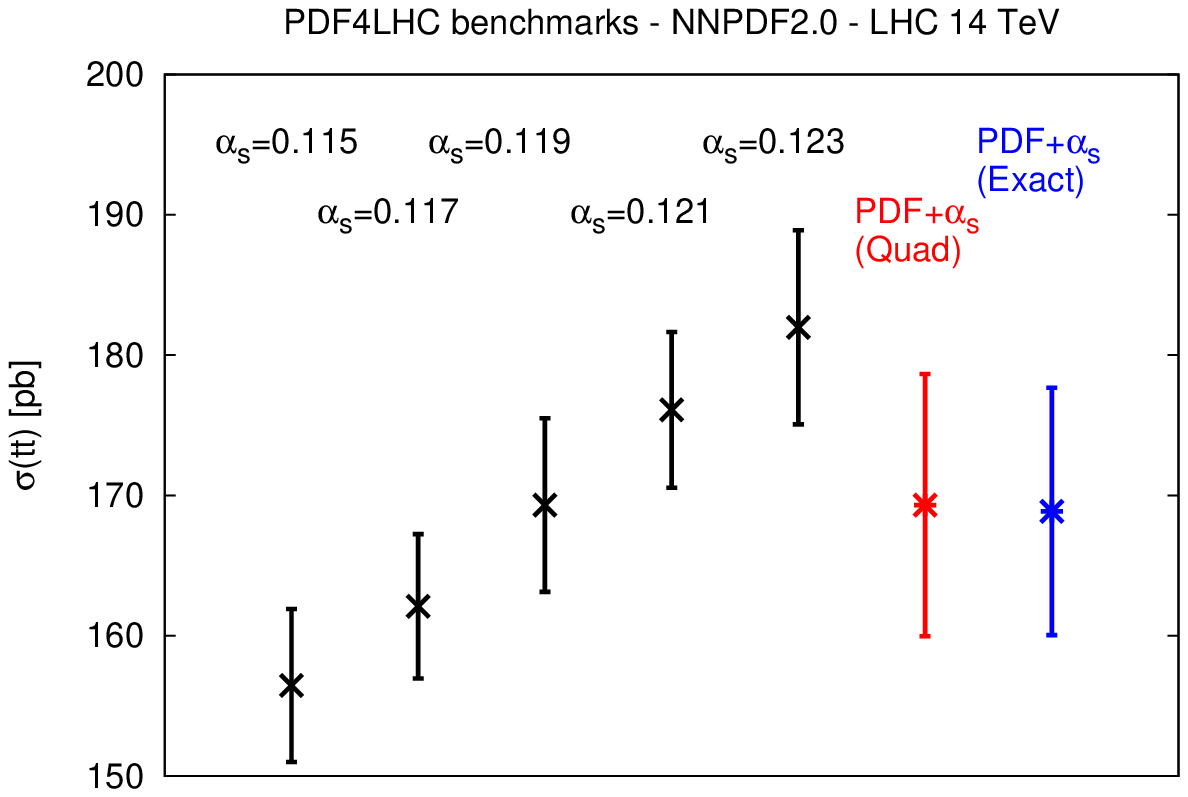}
\epsfig{width=0.45\textwidth,figure=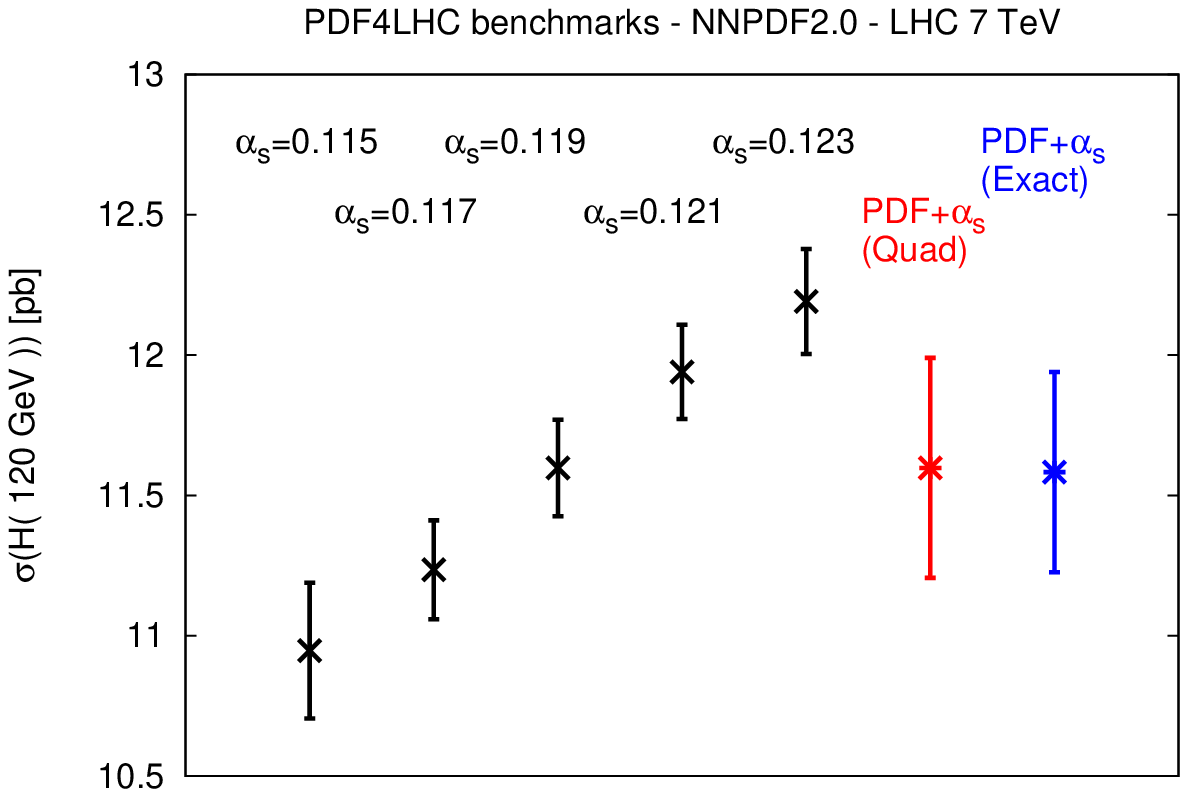}
\caption{\small Predictions for some important LHC observables computed at 7 TeV. 
From top to bottom and from left to right: $W^+$ and
$Z$ production, $t\bar{t}$ production, and
Higgs production in gluon-gluon fusion for $m_H=120$ GeV. Results are
shown for different values of $\alpha_s\lp M_Z\rp$ as well as for the
combined PDF+$\alpha_s$ uncertainties.
 \label{fig:obs7tev}} 
\end{center}
\end{figure}

Results are shown in Fig.~\ref{fig:obs7tev}. 
It is clear that the two methods, quadrature and exact propagation, 
yield essentially identical results. 
Indeed, they ought to give exactly the same result~\cite{Lai:2010nw}
if the combined uncertainty can be obtained as a one--sigma ellipse 
from a quadratic $\chi^2$. 
We also note from Fig.~\ref{fig:obs7tev} that PDF
uncertainties are independent of $\alpha_s$ for any  reasonable 
range of $\alpha_s$. 

For processes which depend on $\alpha_s$ at leading order 
like Higgs or $t\bar{t}$ production, the
combined PDF+$\alpha_s$ uncertainty is as expected sizably 
larger 
than the PDF uncertainty alone: for such processes, comparing
predictions from different PDF sets using a common value of 
$\alpha_s$ is mandatory to obtain a meaningful comparison.

\section{Comparison between global PDF sets}

\begin{figure}[ht]
\begin{center}
\epsfig{width=0.45\textwidth,figure=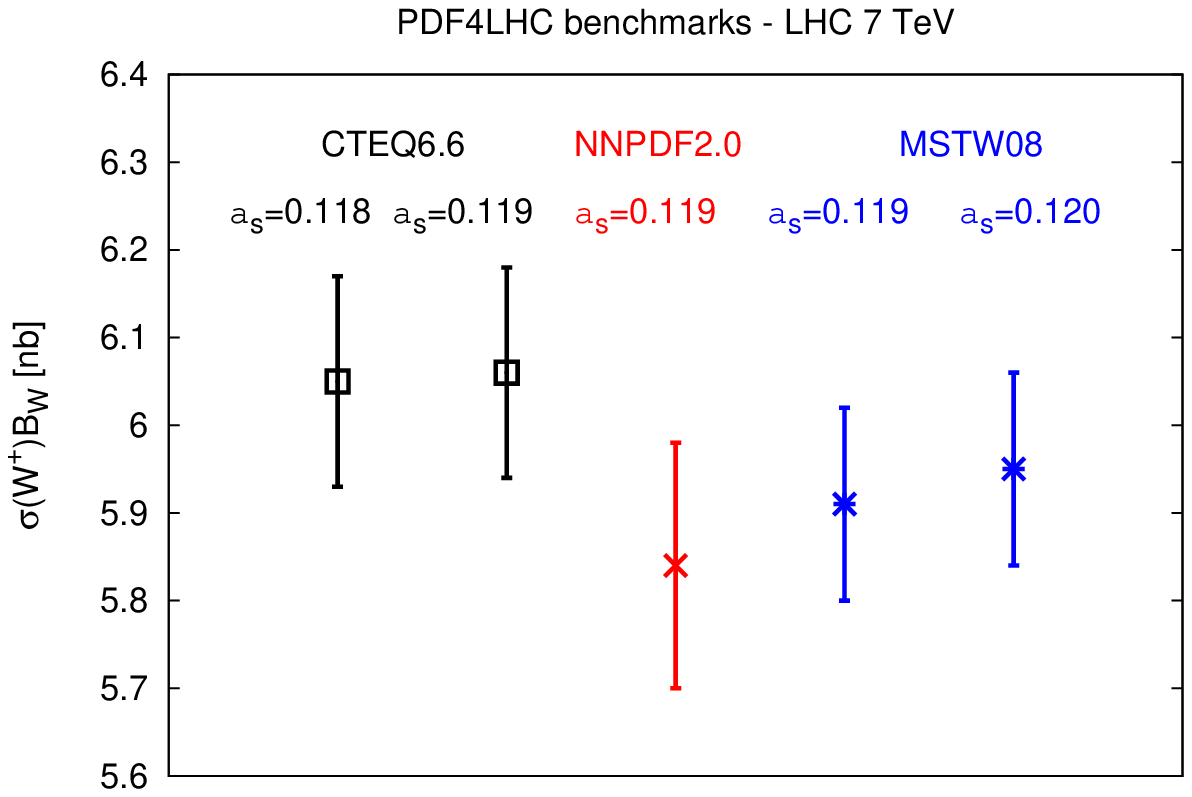}
\epsfig{width=0.45\textwidth,figure=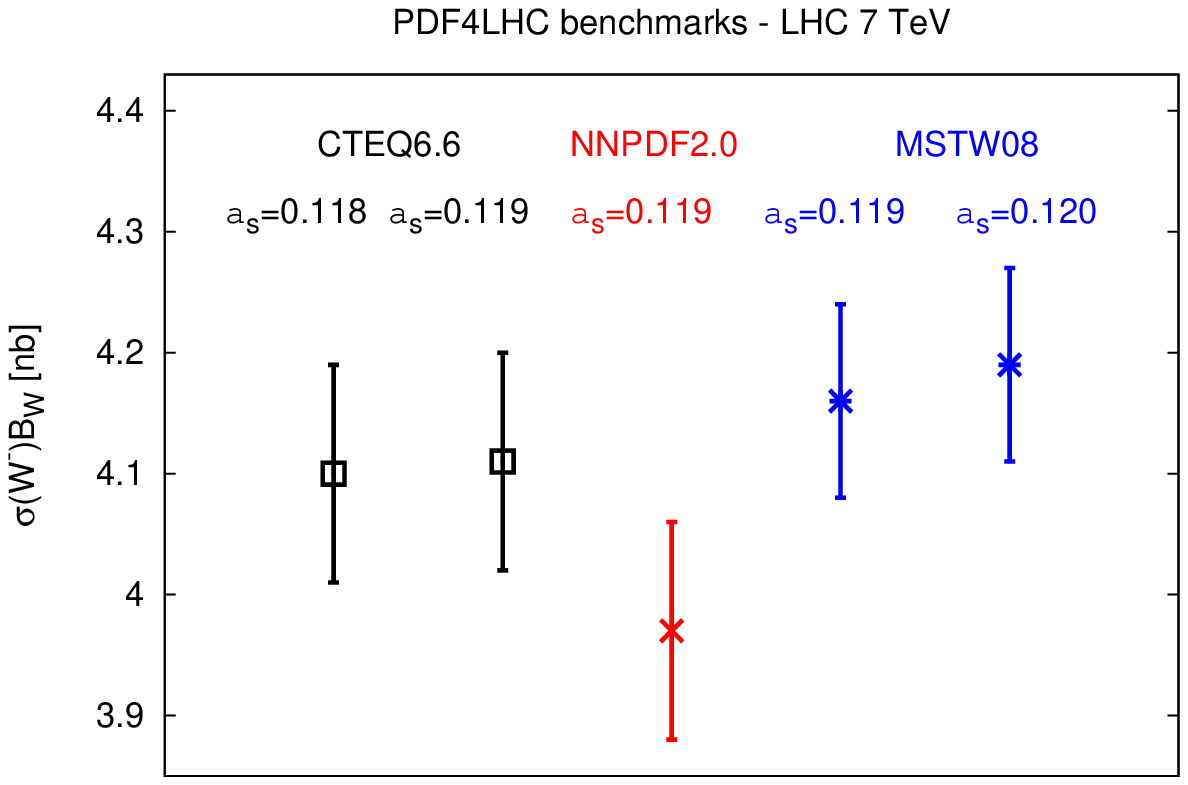}
\epsfig{width=0.45\textwidth,figure=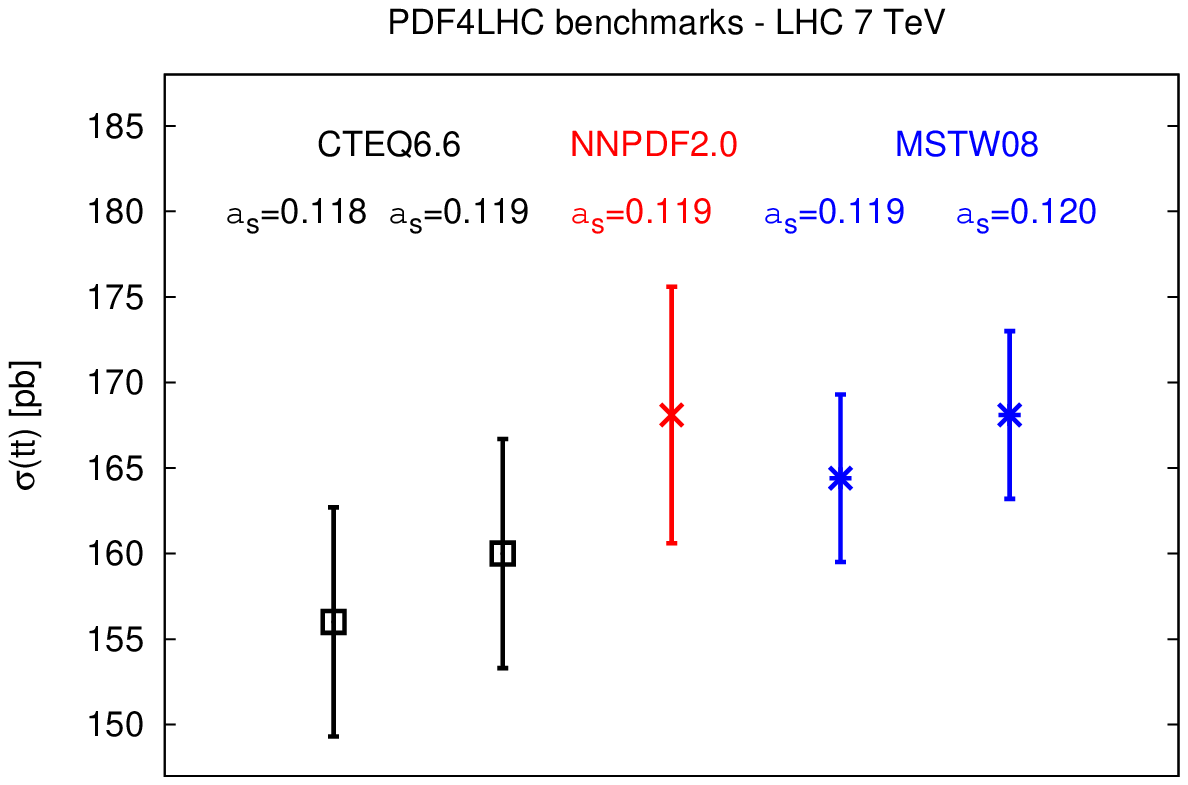}
\epsfig{width=0.45\textwidth,figure=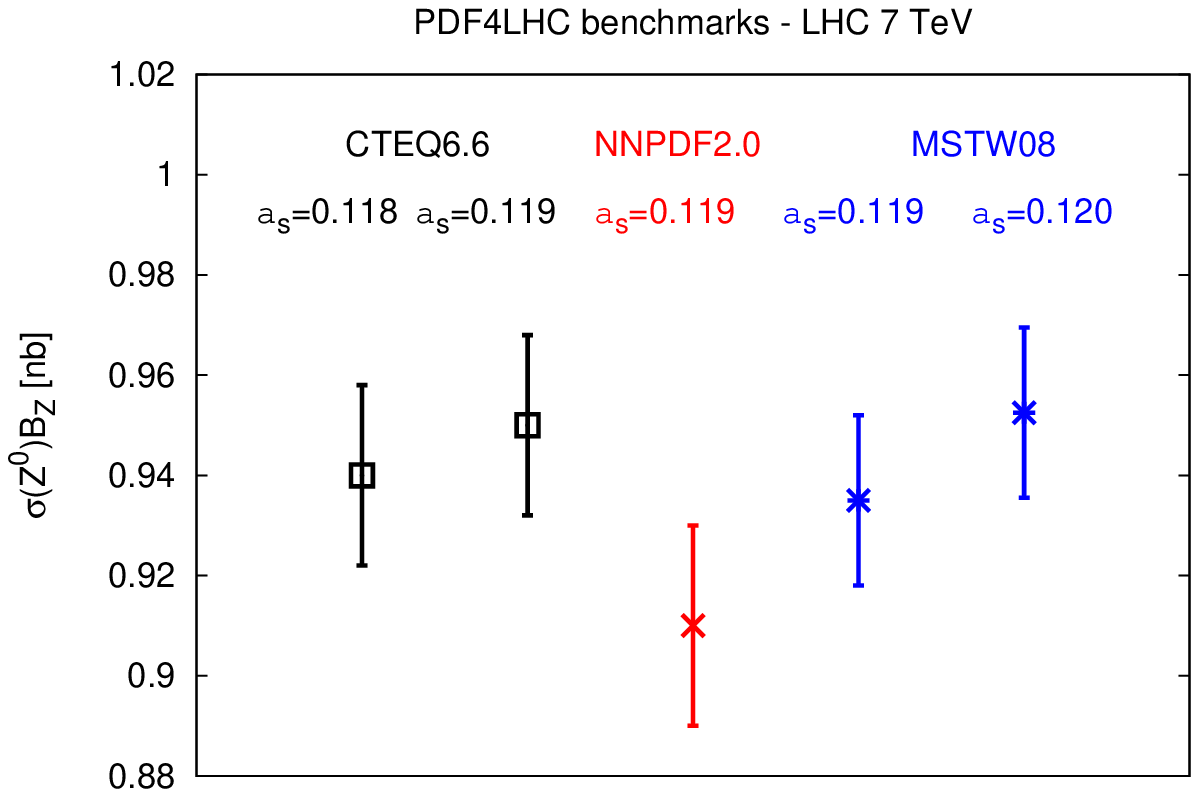}
\epsfig{width=0.45\textwidth,figure=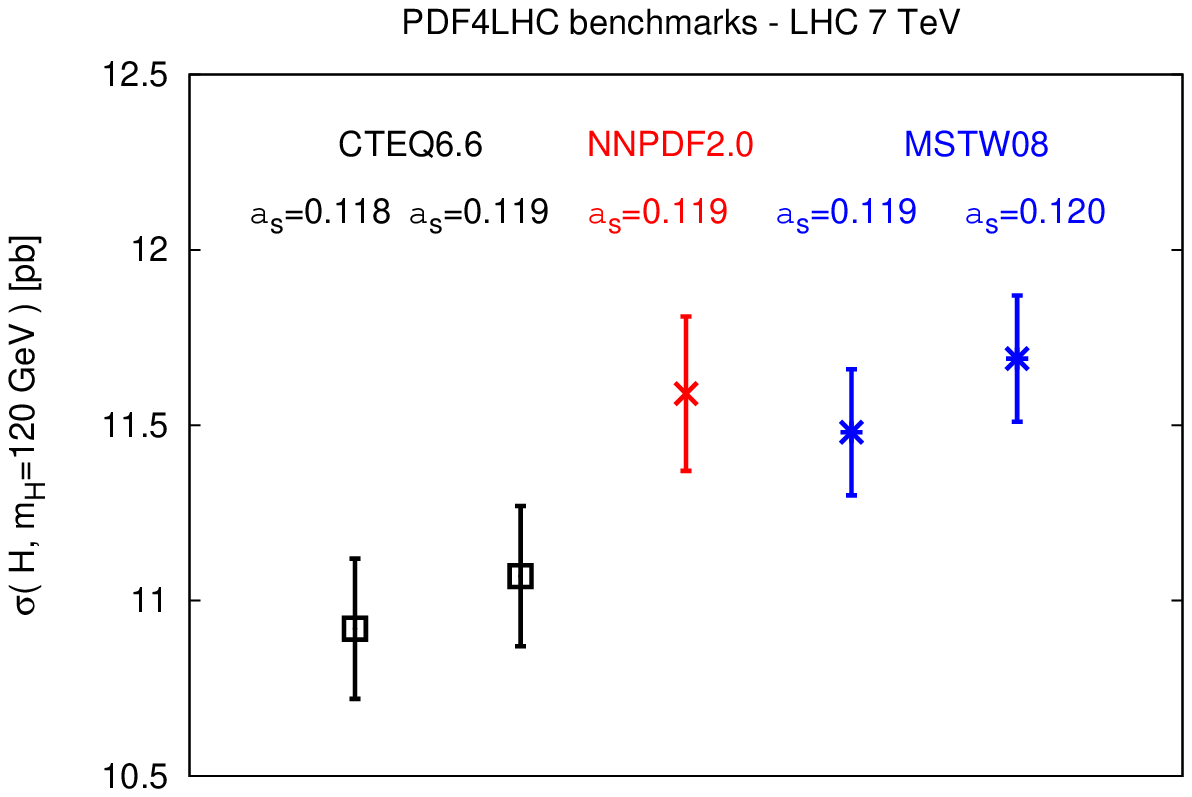}
\caption{\small Comparison
of predictions for LHC observables for NNPDF2.0, MSTW08 and
CTEQ6.6 sets for the LHC at center of mass energy of 7 TeV.
 \label{fig:LHCobs}\label{fig:comp7tev}} 
\end{center}
\end{figure}

Now we compare predictions for important LHC observables
from the three global PDF fits: NNPDF2.0, MSTW08~\cite{Martin:2009iq} 
and CTEQ6.6~\cite{Nadolsky:2008zw}
for the LHC 7 TeV run. The comparison is shown in Fig.~\ref{fig:comp7tev}
and in Table~\ref{tab:LHCobs}. 
For CTEQ and MSTW we show results both at the default value of
$\alpha_s$ and for a common value $\alpha_s\lp M_Z\rp=0.119$. 
 For the
CTEQ6.6 and MSTW08 predictions with $\alpha_s=0.119$
the specific sets from Refs.~\cite{Martin:2009bu,Lai:2010nw}  
have been used. 
We also assume that the PDF uncertainty for these two PDF sets does
not depend in a statistically significant way on the value of $\alpha_s$ when switching from the default to the common value of $\alpha_s$
(which in both cases differ by $\delta\alpha_s=0.001$).
Note that NNPDF2.0
uses as default the value $\alpha_s\lp M_Z\rp=0.119$. 

It is clear 
from Fig.~\ref{fig:comp7tev} that using a common value of the 
strong coupling improves
the agreement between global PDF sets. If predictions with
$\alpha_s=0.119$ are compared, we observe that the three global
PDF sets are in reasonable agreement.
From Table~\ref{tab:LHCobs} is
clear that PDF uncertainties extracted from the NNPDF2.0,
CTEQ6.6 and MSTW08
global fits are quite similar.
We note that a conservative PDF+$\alpha_s$ uncertainty which
accounts for the remaining small discrepancies between PDF sets could
be obtained using the envelope method discussed in 
Ref.~\cite{Demartin:2010er}.

\begin{table}
  \centering
  {\footnotesize
  \begin{tabular}{|c|c|c|c|}
    \hline
        & $\sigma(W^+){\rm Br}\lp W^+ \to l^+\nu_l\rp\,$ [nb]
         & $\sigma(W^-){\rm Br}\lp W^- \to l^+\nu_l\rp\,$ [nb]
         & $\sigma(Z^0){\rm Br}\lp Z^0 \to l^+l^-\rp\,$ [nb]\\
    \hline
    \hline
    NNPDF2.0  &  $5.84\pm 0.14$& $3.97\pm 0.09$ & $0.91 \pm 0.02$ \\
   \hline
   CTEQ6.6 - $\alpha_s=0.118$ & $6.05\pm 0.12$ & $4.10\pm 0.09$ & 
$0.94\pm 0.02$\\
  CTEQ6.6 - $\alpha_s=0.119$ & $6.06\pm 0.12$ & $4.11\pm 0.09$ & $0.95\pm 0.02$ \\
  \hline
   MSTW08 - $\alpha_s=0.119$  & $5.91\pm 0.11$ & $4.16\pm 0.08$ &
$0.94 \pm 0.02$ \\
  MSTW08 - $\alpha_s=0.120$  & $5.95\pm 0.11$ & $4.19\pm 0.08$ &
$0.95 \pm 0.02$ \\
    \hline
  \end{tabular}}\\
\vspace{0.3cm}
 {\footnotesize
  \begin{tabular}{|c|c|c|}
    \hline
        & $\sigma(t\bar{t})$ [pb] 
          & $\sigma(H,m_H=120\,{\rm GeV})$ [pb]
          \\
    \hline
\hline
    NNPDF2.0 & $168.1\pm 7.5$ & $11.59\pm 0.22$ \\
  \hline
   CTEQ6.6 - $\alpha_s=0.118$ & $156.0 \pm 6.7$ & $10.92 \pm 0.20$  \\
  CTEQ6.6 - $\alpha_s=0.119$ &$160.1\pm 6.7$ & $11.07\pm 0.20$ \\
  \hline
   MSTW08 - $\alpha_s=0.119$  & $164.4\pm 4.9$ & $11.48\pm 0.18$ \\ 
  MSTW08 - $\alpha_s=0.120$  & $168.1\pm 4.9$ & $11.69\pm 0.18$ \\
 \hline
  \end{tabular}}
  \caption{\label{tab:LHCobs}  
\small Cross sections for W, Z, $t\bar{t}$ and Higgs production
at the LHC at $\sqrt{s}=7$ TeV and the associated
PDF uncertainties. All quantities have been computed at NLO using
    MCFM for the NNPDF2.0, CTEQ6.6 and MSTW08 PDF sets. All 
uncertainties shown are one--sigma level.
See Fig.~\ref{fig:LHCobs} for the graphical representation
of the results of this table.}
\end{table}

\section{Summary}

We have presented predictions for 
important LHC observables obtained with the NNPDF2.0 set.
We have computed the combined PDF+$\alpha_s$ uncertainties 
on these observables,
and shown that combining errors in quadrature yields an excellent 
approximation to exact error propagation.
The comparison of the NNPDF2.0 results
at the LHC for $\sqrt{s}=$7 TeV with the other global 
PDF analyses, CTEQ6.6 and MSTW08, performed using
a common value of $\alpha_s$ shows a reasonable
agreement both in central values and in uncertainties. 
To understand the remaining moderate
differences between PDF sets a detailed benchmarking
on the lines of the HERA-LHC benchmarks~\cite{Dittmar:2009ii} would
be required.

The NNPDF2.0 PDFs, including sets determined using all
 values of $0.114\le\alpha_s(M_Z)\le 0.124$ in steps
of $\Delta\alpha_s(M_Z)=0.001$, are available from the
NNPDF web site,
\begin{center}
{\bf \url{http://sophia.ecm.ub.es/nnpdf}~.}
\end{center}
They are also available through the LHAPDF interface~\cite{lhapdf}.

\end{document}